\documentclass[fleqn,usenatbib]{mnras}

\usepackage[T1]{fontenc}
\usepackage{ae,aecompl}
\usepackage{color, colortbl}

\usepackage{graphicx}	
\usepackage{amsmath}	
\usepackage{amssymb}	
\usepackage{xcolor}
\usepackage{soul}
\setstcolor{red}
\setul{}{2pt}

\newcommand{\vesc}{\upsilon_\mathrm{esc}}	
\newcommand{\vesco}{\upsilon_\mathrm{esc,0}}	
\newcommand{\vmin}{\upsilon_\mathrm{min}}
\newcommand{\vcirc}{\upsilon_\mathrm{circ}}
\newcommand{\vpp}{\upsilon_\parallel}	
\newcommand{\kpc}{\,\mathrm{kpc}}	

\newcommand{\kms}{\,\mathrm{km}\,\mathrm{s}^{-1}}	
\newcommand{\mvir}{M_\mathrm{vir}}	
\newcommand{\msun}{\rm M_\odot}
\newcommand{\rvir}{r_\mathrm{vir}}	
\newcommand{\g}{\mathrm{G}}	
\newcommand{\bmcmc}{\textsc{bmcmc}}
\newcommand{\LCDM}{$\Lambda{\rm CDM}$} 

\title[M31 escape velocity and dynamical mass modelling]{The Need for Speed: Escape velocity and dynamical mass measurements of the Andromeda galaxy}

\author[P. R. Kafle et al.]
{Prajwal R. Kafle$^{1}$\thanks{E-mail: prajwal.kafle@uwa.edu.au, pkafauthor@gmail.com},
Sanjib Sharma$^{2}$, 
Geraint F. Lewis$^{2}$, 
Aaron S. G. Robotham$^{1}$,
\newauthor
and Simon P. Driver$^{1}$\\\\
$^{1}$ ICRAR, The University of Western Australia, 35 Stirling Highway, Crawley, WA 6009, Australia\\
$^{2}$ Sydney Institute for Astronomy (SIfA), School of Physics A28, The University of Sydney, NSW 2006, Australia}

\date{Accepted 2018 January 8. Received 2018 January 6; in original form 2017 October 17}

\pubyear{2018}

\begin{document}
\label{firstpage}
\pagerange{\pageref{firstpage}--\pageref{lastpage}}
\maketitle

\begin{abstract}
Our nearest large cosmological neighbour, the Andromeda galaxy (M31), is a dynamical system, and an accurate measurement of its total mass is central to our understanding of its assembly history, the life-cycles of its satellite galaxies, and its role in shaping the Local Group environment. 
Here, we apply a novel approach to determine the dynamical mass of M31 using high velocity Planetary Nebulae (PNe), establishing a hierarchical Bayesian model united with a scheme to capture potential outliers and marginalize over tracers unknown distances. 
With this, we derive the escape velocity run of M31 as a function of galacto-centric distance, with both parametric and non-parametric approaches. 
We determine the escape velocity of M31 to be $470\pm{40}\,\kms$ at a galacto-centric distance of $15\,\kpc$, and also, derive the total potential of M31, estimating the virial mass and radius of the galaxy to be  $0.8\pm{0.1}\times10^{12}\,\msun$ and $240\pm{10}\,\kpc$, respectively. 
Our M31 mass is on the low-side of the measured range, this supports the lower expected mass of the M31-Milky Way system from the timing and momentum arguments, satisfying the H\,{\sc i} constraint on circular velocity between $10\lesssim R/\kpc<35$, and agreeing with the stellar mass Tully-Fisher relation. 
To place these results in a broader context, we compare them to the key predictions of the \LCDM\ cosmological paradigm, including the stellar-mass--halo-mass and the dark matter halo concentration--virial mass correlation, and finding it to be an outlier to this relation.
\end{abstract}


\section{Introduction}
Of the many billions of galaxies visible in the ``Observable Universe", only those within the local few megaparsecs are amenable to detailed dissection by our telescopes.
Within this regards, our own Milky Way (MW) and the neighbouring Andromeda (M31), hold pride of place, providing the opportunity to reveal an unparalleled view of the formation and evolution of large galaxies, directly confronting our cosmological paradigm.
Unfortunately, we are buried deep within the MW and thus M31 represents the only large spiral galaxy that can be observed in detail over its entirety, and hence is the focus of numerous observational programs \citep{2004MNRAS.351..117I,2006MNRAS.369..120M,2006ApJ...648..389K,2006MNRAS.369...97H,2009Natur.461...66M,2009ApJ...705.1275G,2012ApJS..200...18D}.

\begin{figure*}
    \centering
	\includegraphics[width=1.6\columnwidth]{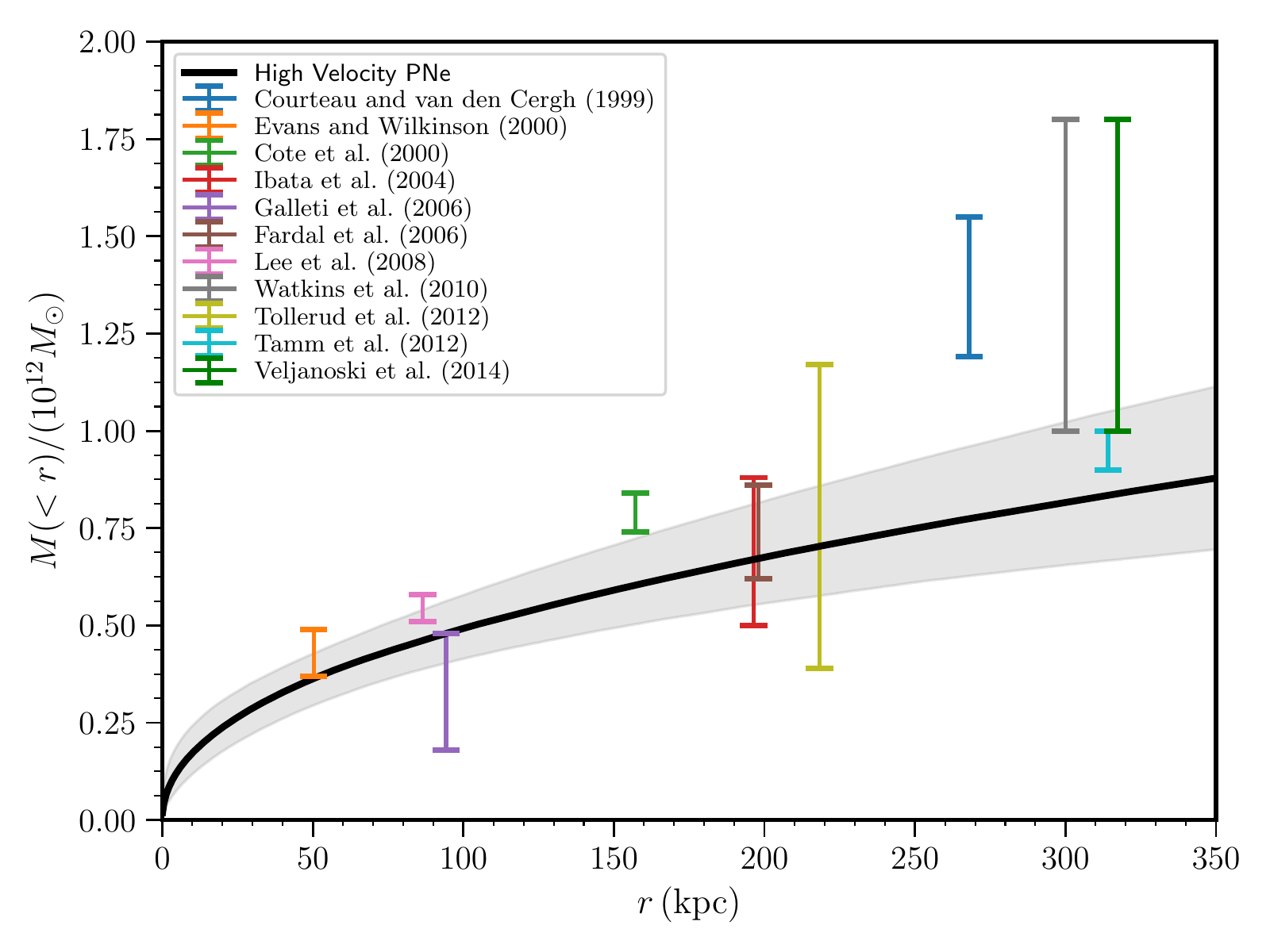}
    \caption{Summary of the recent mass measurements of the M31 galaxy obtained from the different  literature sources (as labelled) that use varieties of techniques. The black solid line with grey shade (showing uncertainties) is the mass profile of the galaxy calculated in this paper by modelling the high velocity Planetary Nebulae (PNe). To convert the projected radius $(R)$ to spherical radius $(r)$, an approximate relation $R = 2\,r/\pi$ is assumed \citep{2006eac..book.....S}.}
    \label{fig:cumumass}
\end{figure*}
An accurate measurement of the mass profile of the M31 galaxy is fundamental in understanding structural and dynamical properties as well as the formation history of the galaxy.
So far several attempts have been made to acquire M31 dynamical mass, which can be classified into  five broad categories:
\begin{enumerate}
\item the rotation curve method \citep{1970ApJ...159..379R,1975ApJ...201..327R,2001ARA&A..39..137S,2006ApJ...641L.109C,2009ApJ...705.1395C},
\item velocity dispersion, virial theorem and tracer mass formalism \citep{1979ApJ...232..699R,1999AJ....118..337C,2000ApJ...537L..91C,2006A&A...456..985G,2008ApJ...674..886L,2010MNRAS.406..264W,2012ApJ...752...45T,2014MNRAS.442.2929V}, 
\item dynamical model or distribution function fitting \citep{2000ApJ...540L...9E,2000MNRAS.316..929E}, 
\item tidal stream orbit modelling \citep{2004MNRAS.351..117I,2013MNRAS.434.2779F}, and 
\item the timing and angular momentum arguments
\citep{2012ApJ...753....8V,2014MNRAS.443.1688D,2014ApJ...793...91G}. 
\end{enumerate}
For an exhaustive review of the topic we refer the reader to \cite{2013MNRAS.434.2779F} while in Fig.~\ref{fig:cumumass} we provide a crude visual summary of the range of the galaxy masses taken from recent literature sources.
The use of different mass tracers, measurements at range of radii, and differences in the approaches to infer the masses make a fair comparison between the reported measurements a daunting task, nonetheless, a convenient summary of all this work is that the total mass of M31 is still uncertain, estimated to be as low as $\sim 0.7\times10^{12}\,\msun$ \citep{2000ApJ...540L...9E,2012A&A...546A...4T} and as high as $\sim2.5\times10^{12}\,\msun$ \citep{2000ApJ...540L...9E,2010MNRAS.406..264W} 
with plethora of measurements in the intermediate range \citep[e.g.][]{2008ApJ...674..886L,2013MNRAS.434.2779F,2014MNRAS.442.2929V}. 

In the light of the huge scatter in the quoted mass of the galaxy, we seek for an alternative way to improve the measurement.  
In this, we attempt an independent measurement of the mass of M31 using the escape velocity inferred from the high velocity tracers, the method first proposed by \cite{1990ApJ...353..486L}.
The method has remained successful in inferring the escape speed and dynamical mass of the MW, of which the studies by \cite{2007MNRAS.379..755S} and \cite{2014A&A...562A..91P} have remained influential.
Both the studies use the same Radial Velocity Experiment \citep[RAVE,][]{2006AJ....132.1645S} survey, albeit different version of the data, of the Galactic disc.
While \cite{2007MNRAS.379..755S} study is only limited to the solar neighbourhood, \cite{2014A&A...562A..91P} improves the method further to explore the radial dependence of the Galactic escape velocity.
More recently, \cite{2017MNRAS.468.2359W} use the halo sample from the Sloan Digital Sky Survey (SDSS) and were able to extend the method out to the MW centric distance of $50\kpc$.

The key reasons for the success of this method are that it is relatively simple and it is empirically powerful as it can estimate the escape velocity from the line-of-sight velocities alone with a similar level of accuracy that can be achieved even when the full phase-space motions are used \citep{1990ApJ...353..486L,1991ARA&A..29..409F}.
The method can be easily adopted for the case of the M31 galaxy or for that matter to any galaxy provided we have enough tracers with high velocity residing in a galaxy.
In this paper, we first improve on the original method by adding additional features such as providing a Bayesian framework, a proper model to capture potential outliers, provisions for the marginalization of the unknown variables and propagation of uncertainties in the observables, a parametric and non-parametric radial fitting for the escape velocities, and eventually use the formalism to make independent predictions for the mass model of M31.

The paper is arranged as follows: In Section~\ref{sec:data}, we introduce our sample of the dynamical tracers of M31, namely Planetary Nebulae (PNe). Section~\ref{sec:hbs} presents Bayesian framework of our modelling scheme. Our results and some model predictions are given in Section~\ref{sec:results}.
In Section~\ref{sec:discussion} we discuss our result to provide it a proper cosmological context. 
Finally, we draw our main conclusions in Section~\ref{sec:conclusion}.
Throughout the paper we denote the normal distribution by $\mathcal{N}$ and the uniform distribution by $\mathcal{U}$. The values for the Hubble constant $H_0= 67.8\,\kms\,\mathrm{Mpc}^{-1}$ and the matter density of the universe $\Omega_m=0.308$ are assumed from the \cite{2016A&A...594A..13P}.

\section{Data}\label{sec:data}

\subsection{Planetary Nebulae (PNe)}\label{sec:pne}
With the development of wide-field multi-object spectrographs \citep{2002PASP..114.1234D} there has been a surge in the number of identified PNe, that happens to be one of the brightest and ubiquitous kinematic tracer of nearby galaxies.
Currently, the PNe catalogue, constructed by \cite{2006MNRAS.369..120M} and \cite{2006MNRAS.369...97H}, comprises the largest publicly available sample of the dynamical tracers of M31, providing sufficiently ample sample with accurate enough line-of-sight velocity information to allow the dynamical modelling of the galaxy. 
The data were observed using the Planetary Nebula Spectrograph on the William Herschel Telescope in La Palma and were made publicly available on-line\footnote{\url{http://www.strw.leidenuniv.nl/pns/PNS_public_web/PN.S_data.html}}.
In total, the published catalogue provides 3300 emission-line objects of which only 2730 are identified as the PNe, and among the likely-PNe sample roughly $6\%$ of the objects are associated with external galaxies and satellite galaxies within M31.
We cull the contaminants and only consider the remaining 2637 PNe that belong to the M31 as our starting sample.
A large fraction of these sources lie in the disk of the galaxy, and are close to the dynamical centre of the galaxy and therefore, they have short orbital time meaning they are relatively better phase-mixed in compare to the tracers in the outskirts of the galaxy where the sub-structures are known to dominate
\citep{2007ApJ...671.1591I,2009MNRAS.396.1842R,2011ApJ...731..124C,2012ApJ...760...76G,2014ApJ...780..128I,2017MNRAS.464.4858K}.

\subsection{Frame of references and coordinate transformations}\label{sec:coord}
To begin with, we assume the central properties of the M31 galaxy as the followings:
\begin{equation*}
\begin{tabular}{lr}
\hline
\hline
\\
Right Ascension & $00^{\text{hh}}42^{\text{mm}}44.33^{\text{ss}}$ \\
Declination & $+41^{\circ}16'07.5''$ \\
Position angle$^1$  & $37.7^{\circ}$\\
Inclination angle$^2$        & $77.5^{\circ}$ \\
Distance from the Sun $(d_\mathrm{M31})^3$  & $780\,\text{kpc}$\\
Heliocentric radial velocity $(v_{\mathrm{M31},\,\mathrm{h}})^4$  & $-301\,\kms$ \\
\\
\hline
\end{tabular}
\end{equation*}
$^1$ \cite{1958ApJ...128..465D}, 
$^2$ \cite{1992ASSL..176.....H},
$^3$ \cite{1998AJ....115.1916H,1998ApJ...503L.131S,2005MNRAS.356..979M,2012ApJ...758...11C}, and
$^4$ \cite{1991rc3..book.....D,2008ApJ...678..187V}.
Additionally, we assume the distance of the Sun from the centre of the MW to be $R_0=8.2\,\kpc$ \citep{2016ARA&A..54..529B}.
In our modelling exercise we require peculiar velocity $(\vpp)$ of dynamical tracers of the M31 galaxy, that is the observed heliocentric line-of-sight velocity of the tracers ($v_\textrm{h}$) with contributions from the solar-reflex $(U_\odot, V_\odot, W_\odot)$, local-standard of rest ($V_\mathrm{LSR}$) and a heliocentric radial motion of the M31 ($\upsilon_{\textrm{M31},\,\mathrm{h}}$) deducted.
This can be done using the following transformation
\begin{equation}\label{eqn:vpec}
\vpp = (\upsilon_{\mathrm{h}} + T)- (v_{\mathrm{M31},\,\mathrm{h}} + T)\cos \Omega,
\end{equation}
with
\begin{equation}\label{eqn:vpecT}
T = U_\odot \cos l \cos b + (V_\odot + V_\mathrm{LSR})\, \sin l \cos b + W_\odot \sin b,
\end{equation}
where $l$ and $b$ are the Galactic longitude and latitude, and $\Omega$ is the angular separation of the tracer from the centre of the M31.
Consistent with the previous dynamical studies \citep[e.g.][]{2000MNRAS.316..929E,2000ApJ...540L...9E,2014MNRAS.442.2929V} we neglect the tangential motion of M31 in our calculation.
This approximation is admissible mainly because the tangential motion of M31 with respect to the MW 
is known to be less than about $17\kms$ \citep{2012ApJ...753....8V} and such small transverse motion would only induce a side-wise motion of a few $\kms$ even for PNe as remote as $\Omega\approx10^{\circ}$,
which is negligible compare to our measurement uncertainties.
Furthermore, to complete the transformation we assume the values for $U_\odot=11.1\,\kms, V_\odot=12.24\,\kms, W_\odot=7.25\,\kms$ and $V_\textrm{LSR}=239.3\,\kms$ \citep{2010MNRAS.403.1829S}. 
Similarly, we desire position vector of the tracer in M31 centric coordinate frame.
Given equatorial coordinates and the heliocentric distances, we can derive respective Cartesian vectors of the M31 as well as its tracers, using say equations B1-B3 of \cite{2007MNRAS.374.1125M}. 
Then to transform the Cartesian position vectors of the M31 tracers from the heliocentric frame of reference (denoted by $x_{\star,\mathrm{h}}$) to the M31 centric frame ($x_{\star, \mathrm{M31}}$), we can shift the position vectors to the centre of M31 and rotate them to the M31 centric coordinate system using 
\begin{equation}\label{eqn:posvec}
\mathbf{x}_{\star,\,\mathrm{M31}} = {\cal R}_{\mathrm{h} \to \mathrm{M31}}(\mathbf{x}_{\star,\,\mathrm{h}} - \mathbf{x}_\mathrm{M31,\,h}),
\end{equation}
where  $\mathbf{x}_\mathrm{M31,\,h}$ is the heliocentric Cartesian vector of the M31 galaxy.  
Here, ${\cal R}_{\mathrm{h} \to \mathrm{M31}}$ is the appropriate rotation matrix and is given by 
\begin{equation}
{\cal R}_{\mathrm{h} \to \mathrm{M31}} = 
\left(
\begin{array}{ccc}
0.770 &  0.324  & 0.549 \\
-0.632 &  0.502  & 0.591 \\
-0.084 &  -0.802 & 0.592
\end{array}
\right),
\end{equation}
which is a function of $R_0$ and central coordinates of the M31 galaxy and is calculated using equation B12 of \cite{2007MNRAS.374.1125M}.

\begin{figure}
    \centering
	\includegraphics[width=1.1\columnwidth]{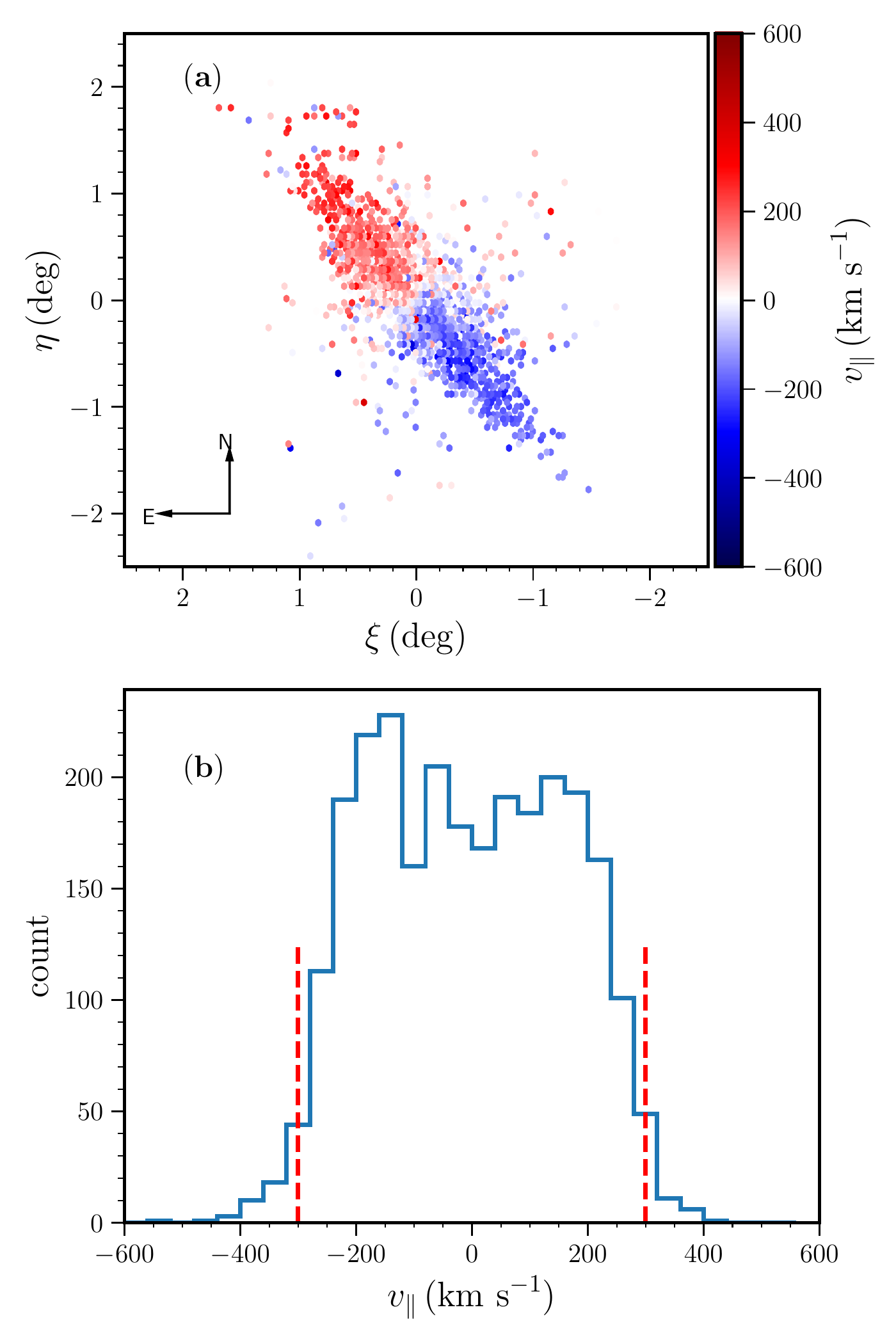}
        \caption{Peculiar velocity distributions of our starting M31 PNe sample, corrected for the M31 systematic motion. (a) Velocity distribution in the conventional on-sky coordinates $\xi$ and $\eta$ in degrees obtained from the gnomonic projection of the tracers Right Ascension and Declination. (b) Distribution of the velocities with red dashed lines showing threshold velocity $\vmin=\pm300\,\kms$ used to classify the high velocity stars $(|\vpp|>|\vmin|)$.}
    \label{fig:vesc_with_vpec}
\end{figure}
In Fig.~\ref{fig:vesc_with_vpec}(a) we show our starting sample of 2637 PNe on-sky projection colour coded by the peculiar velocity ($\vpp$) whereas in (b) we present distribution of $\vpp$.
Here, $\sim85$ stars are seen to be high velocity candidates, depending on the kinematic restriction we apply to classify high velocity stars.
The red vertical dashed lines in panel (b) put at $\pm300\,\kms$ show our default choice of the minimum velocity, discussed later in Section~\ref{sec:hbs}.
The heliocentric distances to these PNe are unfortunately unknown and typical uncertainty in their line-of-sight velocity $\sigma_\upsilon\sim14\,\kms$ \citep{2006MNRAS.369..120M}.

\section{Method: Hierarchical Bayesian Modelling and Accessories}\label{sec:hbs}
Below we present an outline of the method that comprises three major steps of our approach: determination of the escape velocity, simple model predictions (derivation of cumulative mass, potential and circular velocity runs) and inference of the virial properties.

\subsection{Escape velocity curve modelling}
\subsubsection{Setting up the inference}\label{sec:modelsetup}
Similar to \cite{1990ApJ...353..486L}, we start by assuming that the high velocity $\{\vpp: |\vpp|>\vmin \}$ wings of the distribution function of the galaxy follow a power-law distribution and can be expressed as
\begin{equation}\label{eqn:df}
f(\vpp|r) = f_0\,(v_\textrm{esc}(r) - |\vpp|)^{k+1},
\end{equation}
where the normalization, $f_0$, is given by
\begin{equation}\label{eqn:fnorm}
\begin{aligned}
f_0 &= \int_{\vmin}^{\vesc} f(\vpp|r)\,\text{d}\,\vpp\\
& = \frac{k+2}{(\vesc(r) - \vmin)^{k+2}}.
\end{aligned}
\end{equation}
The $\vmin$ is a threshold velocity, that is the tracer with the absolute peculiar velocity ($|\vpp|$ in equation~\ref{eqn:vpec}) larger than $\vmin$ is classified as a high velocity tracer and makes it to our final sample.
Generally, in the literature $\vmin$ is assumed to be $300\,\kms$ \citep{1990ApJ...353..486L,2007MNRAS.379..755S,2014A&A...562A..91P}
and we also make it our default choice, however later we investigate the effect of our choice in final results. 
Similar to \cite{2014A&A...562A..91P} we express the escape speed as a function of the tracer galacto-centric distance $r$, this is to allow us to infer the spatial run of the speed.
The specific details about the exact expressions for the $\vesc(r)$ are provided in Section~\ref{sec:escvel}.

Some tracers in our data could probably be outliers (possibly unbound stars) and to capture this we adopt an outlier, or background model, $g$, which is given by
\begin{equation}\label{eqn:outlier}
g(\vpp) = g_0 \mathcal{T}(\vpp|0, \sigma_g, \nu_g),
\end{equation}
and the model is normalized between $[\vmin, \infty]$ such that
\[g_0 = \frac{1}{2} - \frac{1}{\pi} \tan^{-1} (\vmin/\sigma_g).\]
Here, ${\cal T}$ represents the Student t-distribution, defined by three parameters mean, variance $(\sigma_g^2)$ and degree of freedom $(\nu_g)$, and is known to approximate the ${\cal N}$ distribution as $\nu_g \to \infty$.
The ${\cal T}$ distribution is preferable over the ${\cal N}$ distribution as the former has broader wings.
We fix $\sigma_g=1000 \kms$ and $\nu_g = 1$, but investigate the effects of these assumptions in our analysis in Section~\ref{sec:results}.
Additionally, we like to put a note here that instead of introducing an outlier model and thereby increasing the modelling complexity, one could simply use $3\sigma$ or $5\sigma$ velocity clippings to get rid of outliers at the earlier data curating step.
However, given the potential future application of the modelling scheme, as new data arrive, it is therefore sensible to adopt the former, a comparatively superior approach.

Equipped with the above models, now we can express a joint probability distribution for the `complete-data' (high velocity and outlier stars) as
\begin{equation}\label{eqn:likeli}
p(\vpp^t|r) = \eta\,f(\vpp^t|r) + (1-\eta)\, g(\vpp^t),
\end{equation}
where $\eta$ indicates the outlier fraction, $\vpp^t$ represents the clean, error-free `true' version of the observed peculiar velocity ($\vpp$) and $\sigma_\upsilon$ is the error in $\vpp$.
Since $\vpp^t$ is not directly observed, we treat it as a latent variable and assume the errors $\sigma_v$ are Gaussian $({\cal N})$. 
Finally, we can construct the full likelihood function as
\begin{equation}\label{eqn:flikeli}
\mathcal{L}(\Theta) = p(\vpp^t|\Theta, r)\,p(r|\alpha)\,\mathcal{N}(\vpp|\vpp^t, \sigma_\upsilon),\\    
\end{equation}
\begin{figure}
    \centering
	\includegraphics[width=1\columnwidth]{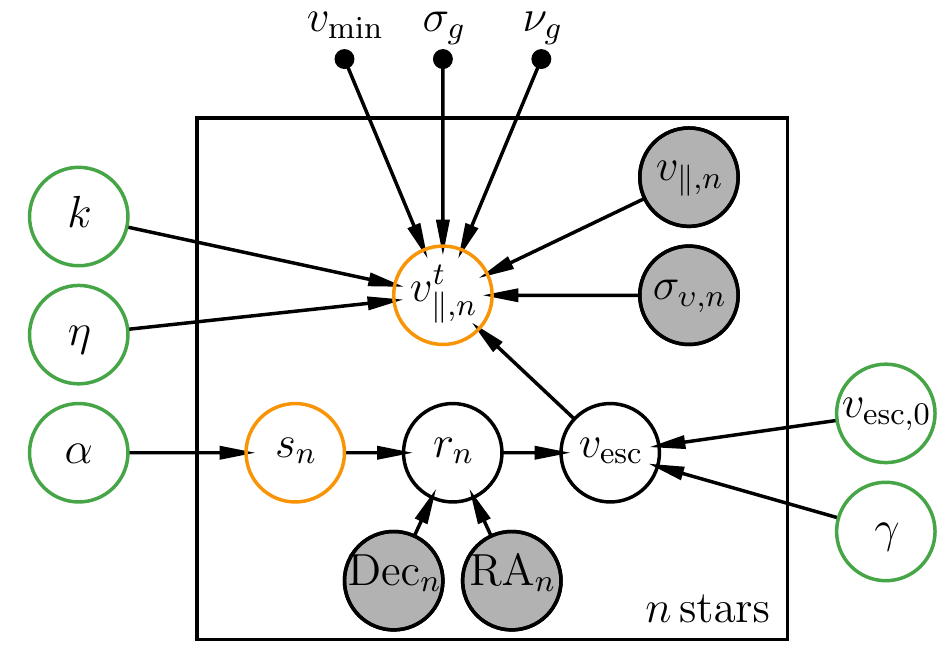}
    \caption{A plate diagram for escape velocity modelling. It shows a graphical illustration of the dependencies between the observed data (grey shaded circle) and model parameters that are kept free (green circle) or fixed (black dot) in the form of a probabilistic graphical model. Orange circles represent the latent parameters.}
    \label{fig:platediag}
\end{figure}
Here, $p(r|\alpha)$ is the tracer density profile, which for simplicity we assume to be a single power-law of the galacto-centric distance $r$, that is 
\begin{equation}\label{eqn:density}
p(r|\alpha) \propto r(\mathbf{x}_{\star,\,\mathrm{M31}})^{-\alpha}.
\end{equation}
The distance, $r$, is a function of the position vector, $\mathbf{x}_{\star,\,\mathrm{M31}}$, that is related to the observables $\mathrm{RA}$ and $\mathrm{Dec}$, and unknown heliocentric distances, $s$, of the tracers via equation~\ref{eqn:posvec}.
As we do not know the distance $s$ to the tracers we treat it as a latent parameter. 
Furthermore, we keep the logarithmic power-law slope $\alpha$ of the tracer density distribution free in the range between 2 and 3.5, 
in line with recent literature \citep[e.g.][]{2005ApJ...628L.105I,2007ApJ...671.1591I,2011ApJ...739...20C,2012ApJ...760...76G,2014ApJ...780..128I}. 

\subsubsection{Expressions of Escape Velocity}\label{sec:escvel}
We can determine $\vesc$, required for equation ~\ref{eqn:df}, either parametrically, or non-parametrically.
In the first approach, we express $\vesc$ in a simple parametric form given by
\begin{equation}\label{eqn:vescp}
\vesc(r) = \vesco \,(r/r_s)^{-\gamma},
\end{equation}
with the scale-length, $r_s$, fixed at $15\kpc$.

In the other approach, we assume that the $\vesc(r)$ profile over the radial extent of the data is a linear interpolation of $\upsilon_{\mathrm{esc},j}$ measured at distance $r_j$, where the number of nodes $j$ and corresponding radius $r_j$ are pre-specified.
We can express this form for $\vesc$ as the following
\begin{equation}\label{eqn:vescnp}
\vesc(r) = \mathrm{Interpolate}\,(r_j, \upsilon_{\mathrm{esc},\,j}). 
\end{equation}
Here, the number of nodes $j$ can be viewed as a parameter that provides the radial resolution of $\vesc(r)$.
Ideally, $j$ can be kept infinitely large to fully capture the $\vesc$ range of the galaxy. 
However, for practical purposes we keep $j=6$ and calculate $\{\upsilon_{\mathrm{esc},1},....,\upsilon_{\mathrm{esc},6}\}$ at $r/\kpc=5,10,15,20,25\,\mathrm{and}\,30$; 
essentially $\vesc(r)$ that enters into equation~\ref{eqn:df} is linear interpolation of $\upsilon_{\mathrm{esc},j}$ at the specified radii $r_j$. 

The probabilistic graphical model indicating the interdependences of our model parameters and observables are shown in  Fig.~\ref{fig:platediag}.
Syntax-wise, in the case of the non-parametric approach we replace $\{ \gamma, \vesco \}$ in $\Theta$ with $\{\upsilon_{\mathrm{esc},1},..\upsilon_{\mathrm{esc},j}\}$ while the rest of the steps remain exactly same.

\subsubsection{Prior and posterior distributions}
Ultimately, in the Bayesian format the posterior distribution that we like to sample can be written as the product of priors and likelihood function as the following
\begin{equation}\label{eqn:posterior}
p(\Theta, s_t, \vpp^t|\vpp) \propto p(k,\vesco)\, \prod_n \mathcal{L}(\Theta),
\end{equation}
where product is over all $n$ stars.
Here, $\Theta$ represents the set of model parameters  $\{ k, \vesco, \gamma, \alpha, \eta \}$ that we aim to constrain, which forms the higher level of our hierarchical model whereas the heliocentric distance $s$ and the error free peculiar velocity $\vpp^t$ of the tracers make the lower level of our model.
Here $p(k,\vesco)$ represent priors on $k$ and $\vesco$ parameters.
For different Milky-Way like simulated galaxies obtained from cosmological simulations the maximum value of $k$ is found to cover a range between 2.3-4.7 \citep{2014A&A...562A..91P}.
Following this we consider $p(k)\in{\cal U}[0,4.7]$ as a prior on $k$.
Similarly, for $\vesco$ also we adopt the (less) informed uniform prior, $p(\upsilon_\mathrm{esc,\{j: j \in R\}}/\kms) \in{\cal U}[\vmin,1000]$.
Also, we assume flat priors for the remaining parameters, namely, $\eta\in{\cal U}[0,1]$ and $\gamma\in{\cal U}[0,1]$.
The heliocentric distance $s/\kpc$ of each tracer is proposed between ${\cal U}[600,1100]$ 
constrained to follow the density distribution given by equation~\ref{eqn:density} with hyper-prior $\alpha\in{\cal U}[2,3.5]$ and the proposal distribution for $|\vpp^t|/\kms$ is assumed to follow ${\cal U}[\vmin, |\vpp| + 5\sigma_\upsilon]$.

\subsection{Mass modelling}\label{sec:massmodel}
With the escape velocity curve next we can derive the mass profile and the virial properties of the galaxy.
For this we interpret the escape velocity, $\vesc$, as the minimum speed required to reach some limiting radius, $r_\mathrm{max}$, and relate it to the galactic potential using the relation
\begin{equation}\label{eqn:pot}
\vesc = \sqrt{2\,|\Phi(r) -\Phi(r_\mathrm{max})|}, 
\end{equation}
where $\Phi$ represents the total potential of the galaxy at a given radius $r$.
Similar to \cite{2007MNRAS.379..755S} and \cite{2014A&A...562A..91P}, we also assume $r_\mathrm{max}$ to be thrice that of the virial radius, a sufficiently large distance.
Since the gravitational potential is a weak function of radius at such large distances, we expect results are insensitive to small changes in the limiting radius, which is demonstrated to be a valid assumption in \cite{2014A&A...562A..91P}.
The above equation can be solved to express $\Phi(r)$ as a function of $\vesc$, which can then be fitted to a mass model of the M31 and the model can be finally used to infer key dynamical properties of the galaxy.
But first, we must construct a realistic model of the galaxy.

We assume that the two most dominant stellar components of the M31 that is bulge and disk follow \cite{1990ApJ...356..359H} and \cite{1975PASJ...27..533M} type potentials given by  
\begin{equation}\label{eqn:bulge}
\Phi_\mathrm{b} = -\frac{\g\, M_\mathrm{b}}{r + q}\,
\end{equation}
and
\begin{equation}\label{eqn:disk}
\Phi_\mathrm{d} = -\frac{\g\, M_\mathrm{d}}{\sqrt{R^2 + \left( a + \sqrt{z^2 + b^2} \right) ^2 }},
\end{equation}
respectively, where $M_\mathrm{b}$ and $q$ represent the bulge mass and scale-length, whereas $M_\mathrm{d}$, $a$ and $b$ are for the disk mass, scale-length and scale-height respectively. 

In addition to the bulge and disk models, we assume the potential of the dark matter halo to have a Navarro-Frenk-White \cite[NFW,][]{1996ApJ...462..563N} profile given by
\begin{equation}
 \Phi_\mathrm{h} = \frac{-\g\,\mvir \ln(1+r\,c/\rvir)}{g(c)\,r},
\end{equation}
with $g(c) = \ln(1+c) - c/(1+c)$, and 
\begin{equation}\label{eqn:rvir}
\mvir  = \frac{4\,\pi}{3} \rvir^3 \Delta\, \rho_c\, 
\end{equation}
where $\mvir$, $\rvir$ and $c$ denote the virial mass, virial radius and concentration parameter,
whereas $\Delta$ is the virial over-density parameter and $\rho_c = 3H_0^2/(8 \pi G)$ is the critical density of the Universe.
Generally, two contrasting choices for $\Delta$ are preferred in the literature. 
First, for the spherical collapse definition $\Delta = \Omega_m \delta_\mathrm{th}$, 
where  $\delta_\mathrm{th}=340$ is an over-density of the dark matter compared to the average matter density \citep{1998ApJ...495...80B}, that is $\Delta\simeq100$.
We denote virial mass and radius for this cases as $\mvir$ and $\rvir$ respectively.
Second, some literature assume $\Delta=200$, and in this case we label the corresponding mass and radius as $M_{200}$ and $r_{200}$ respectively.
The different choices for $\Delta$ is just a matter of convention and there is no correct or incorrect choice.
The outer radius and mass enclosed are inversely proportional to the $\Delta$ parameter as for a system with monotonically decreasing density distribution 
200 times over-density would occur at smaller radius than 100 times over-density compare to the average matter density of the universe. 
We report final measurements for both the cases. 

Finally, we combine the contribution of each component of the galaxy and express the total potential of the galaxy as
\begin{equation}\label{eqn:ptot}
\Phi_\mathrm{t}(r) = \Phi_\mathrm{b}(r) + \langle \Phi_\mathrm{d} (R, z) \rangle_\theta + \Phi_\mathrm{h}(r).
\end{equation}
Clearly, the bulge and halo models are spherically symmetric but the disk model is a function of both cylindrical radius ($R = r \sin \theta$) and height above the plane of the disk ($z = r \cos \theta$),
and is azimuthally averaged.

There are total of seven parameters ($M_{\mathrm{b}}$, $q$, $M_{\mathrm{d}}$, $a$, $b$, $\mvir$, $c$) that completes our bulge-disk-halo model, of which the parameters of the baryonic components are assumed as the following
\begin{equation}\label{eqn:bdparams}
\begin{tabular}{lr}
\\
Bulge scale-length ($q$) & $0.7\,\kpc$, \\
Bulge mass ($M_{\mathrm{b}}$) & $3.4\times 10^{10}\,\msun$, \\
Disk scale-length ($a$) & $6.5\,\kpc$, \\
Disk scale-height ($b$) & $0.26\,\kpc$, and \\
Disk mass ($M_{\mathrm{d}}$) & $6.9\times10^{10}\,\msun$.\\\\
\end{tabular}
\end{equation}
The values for scale-lengths and scale-height are adopted from \cite{2001ApJ...557L..39B} and \cite{2006AJ....131.1436F} whereas the masses (in the units of $10^{10}\msun$) are taken to be a straight average of the most recent estimates obtained from 
\citet[][$M_{\mathrm{b}}=3.2,M_{\mathrm{d}}=7.2$, a case with the disk mass-to-light ratio of 3.3]{2006MNRAS.366..996G}, 
\citet[][$M_{\mathrm{b}}=3.5,M_{\mathrm{d}}=5.8$, inferred from the \emph{Spitzer} 3.6-micron imaging data and $B-R$ color profile]{2008MNRAS.389.1911S}, 
\citet[][$M_{\mathrm{b}}=2.32,M_{\mathrm{d}}=7.1$, bulge mass is derived from stellar velocity dispersions and the disk mass is determined from stellar population models]{2009ApJ...705.1395C}, \citet[][$M_{\mathrm{b}}=3.8,M_{\mathrm{d}}=8.8$, estimated by fitting bulge, disc, gas and dark matter halo combined model to H\,{\sc i} kinematics data]{2010A&A...511A..89C} and \citet[][$M_{\mathrm{b}}=4.4,M_{\mathrm{d}}=5.7$, derived from fitting the rotation curve]{2012A&A...546A...4T}.
The remaining two parameters, $c$ and $\mvir$, that defines the dark matter halo are kept free and constrained via the maximum likelihood analysis.
The likelihood function or the total probability of obtaining the data (total potential $(\Phi_t)$ given the model parameters ($c$ and $\mvir$) that we aim to maximise is given by
\begin{equation}\label{eqn:lnlikelipot}
p(c, \mvir|\Phi(r)) = \prod_j {\cal N}(\Phi_t(r_j)|\Phi(r_j), \sigma_{\Phi(r_j)}).
\end{equation}

\begin{figure*}
    \centering
	\includegraphics[width=1.8\columnwidth]{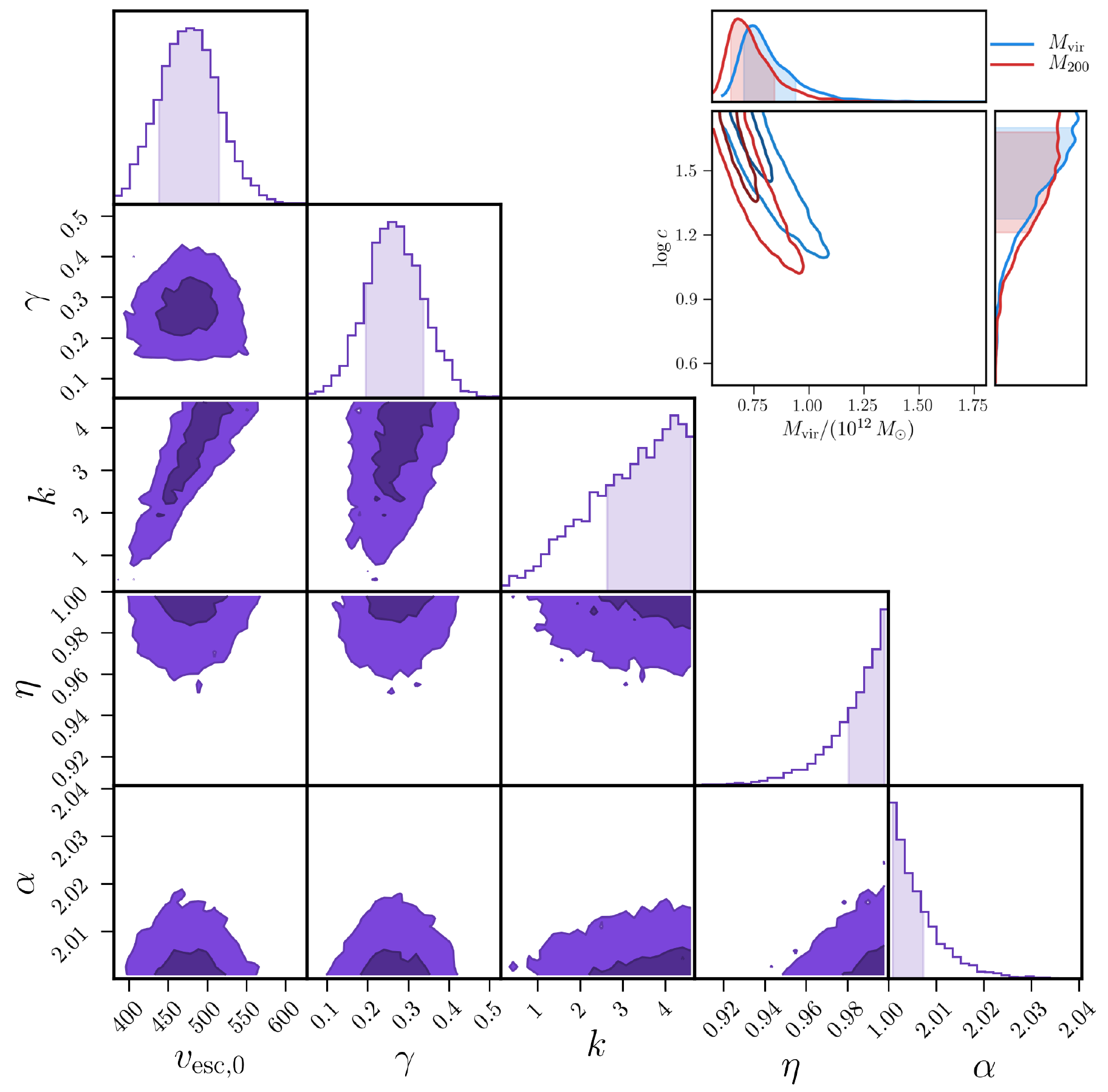}
    \caption{Joint posterior distributions of the model parameters obtained from the escape velocity and mass modellings, showing the default case, where $\vmin=300\,\kms$, $r_s=15\,\kpc$, $d_\mathrm{M31}=780\kpc$, $v_\mathrm{M31,h}=-301\kms$, $\sigma_g=1000\,\kms$ and $\nu_g=1$ are assumed. Panel in the inset shows the joint distributions of concentration and virial mass obtained from the mass modelling for the corresponding case, where blue and red distributions show results for two different definitions of the halo masses as labelled.}
    \label{fig:posteriors}
\end{figure*}

Having established the Bayesian framework for our problem, we like to ultimately sample the posterior distribution $p(\Theta, s_t, \vpp^t|\vpp)$ (equation~\ref{eqn:posterior}) and $p(c, \mvir|\Phi(r))$ (equation~\ref{eqn:lnlikelipot}).
Below we briefly discuss the sampling technique.

\subsection{Inference: Metropolis within Gibbs (MWG)}
The posterior distribution $p(\Theta, s_t, \vpp^t|\vpp)$ is a function of the set of model parameters $\Theta=\{ \vesco, \gamma, \eta, \alpha, k \}$ or $\Theta=\{ \{ \vesco ..., v_\mathrm{esc,j} \}, \gamma, \eta, \alpha, k \}$ depending on whether we adopt the parametric (equation~\ref{eqn:vescp}) or the non-parametric (equation~\ref{eqn:vescnp}) form for $\vesc$.
To sample the posterior we make use of the Metropolis within Gibbs sampler from \bmcmc\ software\footnote{\url{https://github.com/sanjibs/bmcmc}} \citep[][]{2017ascl.soft09009S,2017ARA&A..55..213S}.
Firstly, using the general Metropolis-Hastings algorithm, a set of values for the model parameters, $\Theta$, is proposed from a proposal distribution.
Note the posterior distribution function also explicitly depends on the known unknowns or the latent variables such as the heliocentric distance and true peculiar velocity, $\vpp^t$, of the tracers.
We must marginalize the distribution function over these latent variables before we can infer anything about the model parameters.
Unfortunately, the expression for the likelihood function given in Equation~\ref{eqn:flikeli} is non-analytical meaning marginalization must be be done numerically.
Therefore, we set-up a multi-level (hierarchical) inference where we could directly consider the latent variables as hyper-parameters and sample them for each proposal of the model parameters.
For this we employ the \bmcmc\ that uses Gibbs sampling method for the marginal inference.
In short, given a set of observables, $(\alpha, \delta, {\cal R}, \vpp, \sigma_\upsilon)$, for each tracer along with the proposed values for $(\vpp^t, s)$ and assuming a galacto-centric number density distribution (equation~\ref{eqn:density}), \bmcmc\ yields the posterior distributions in the form of walks/chains of the model parameters $\Theta$ and ancillary data such as the log-likelihood value at each walk.
We use these chains for the final inference, and also where needed we propagate them to derive physical quantities of the interest.

Similarly, in the separate case of fitting the mass model via equation~\ref{eqn:lnlikelipot}, we again make use of the \bmcmc. 
In this case there is no latent variable and hence, \bmcmc\ only uses the general Metropolis Hastings algorithm to sample the distribution for the model parameters $\{\mvir, c\}$.
For all the undertaken experiments, we run MWG for sufficient autocorrelation time to ensure that the distributions of parameters are stabilized around certain values.
We report the median of the probability distribution of a model parameter as its best estimate and $16^{\rm th}$ and $84^{\rm th}$ percentiles of the distribution as measure of its uncertainty.

\section{Results}\label{sec:results}
\begin{table*}
\caption{Estimated model parameters from different analysis. When unspecified, the default values for the model parameters are assumed, i.e., $\vmin=300\,\kms$,  $r_s=15\,\kpc$, $d_\mathrm{M31}=780\kpc$, $\upsilon_{\mathrm{M31,h}}=-301\,\kms$, $\sigma_g=1000\,\kms$ and $\nu_g=1$.}
\label{table:pneresult}
\begin{tabular}{@{}lccccccc}
\hline\\
Cases & Number of& $\vesco$ & $\gamma$ & $\eta$ & $k$ & $\mvir(M_{200})$ & $\rvir(r_{200})$ \\ 
  & PNe    &(km\,s$^{-1}$) &  &  &  &$(10^{12}M_\odot)$ & $(\kpc)$\\ 
\hline\\
\rowcolor{lightgray} Default & 86 & $470^{+40}_{-40}$ & $0.26^{+0.07}_{-0.07}$ & $0.99^{+0.02}_{-0.01}$ & $3^{+1}_{-1}$ & $0.8^{+0.1}_{-0.1}$($0.7^{+0.1}_{-0.1}$) & $240^{+10}_{-10}$($188^{+7}_{-11}$) \\  \\ 
$d_{\mathrm{M31}} = 790\,\kpc$ & 86 & $480^{+40}_{-40}$ & $0.27^{+0.07}_{-0.07}$ & $0.99^{+0.02}_{-0.01}$ & $3^{+1}_{-1}$ & $0.8^{+0.1}_{-0.1}$($0.7^{+0.1}_{-0.1}$) & $240^{+10}_{-10}$($188^{+7}_{-11}$) \\  \\ 
$d_{\mathrm{M31}} = 770\,\kpc$ & 86 & $470^{+40}_{-40}$ & $0.27^{+0.07}_{-0.07}$ & $0.99^{+0.02}_{-0.01}$ & $3^{+1}_{-1}$ & $0.8^{+0.1}_{-0.1}$($0.7^{+0.07}_{-0.12}$) & $240^{+10}_{-10}$($187^{+7}_{-10}$) \\  \\ 
\rowcolor{lightgray} $\upsilon_{\mathrm{M31,h}} = -296\,\kms$ & 87 & $490^{+40}_{-40}$ & $0.25^{+0.07}_{-0.07}$ & $0.99^{+0.02}_{-0.01}$ & $3^{+1}_{-1}$ & $0.9^{+0.1}_{-0.2}$($0.8^{+0.1}_{-0.2}$) & $250^{+10}_{-20}$($190^{+10}_{-10}$) \\  \\ 
\rowcolor{lightgray} $\upsilon_{\mathrm{M31,h}} = -306\,\kms$ & 90 & $470^{+40}_{-40}$ & $0.27^{+0.07}_{-0.07}$ & $0.99^{+0.02}_{-0.01}$ & $3^{+1}_{-1}$ & $0.73^{+0.07}_{-0.13}$($0.66^{+0.06}_{-0.11}$) & $240^{+10}_{-10}$($184^{+6}_{-9}$) \\  \\ 
$\vmin=290\,\kms$ & 111 & $470^{+40}_{-30}$ & $0.27^{+0.06}_{-0.06}$ & $0.99^{+0.01}_{-0.01}$ & $3^{+1}_{-1}$ & $0.76^{+0.07}_{-0.12}$($0.69^{+0.06}_{-0.11}$) & $239^{+7}_{-12}$($186^{+6}_{-9}$) \\  \\ 
$\vmin=310\,\kms$ & 65 & $480^{+40}_{-40}$ & $0.24^{+0.08}_{-0.08}$ & $0.98^{+0.03}_{-0.01}$ & $3^{+1}_{-1}$ & $0.9^{+0.1}_{-0.2}$($0.8^{+0.1}_{-0.2}$) & $250^{+10}_{-20}$($190^{+10}_{-10}$) \\  \\ 
\rowcolor{lightgray} $r_{s}=14\,\kpc$ & 86 & $480^{+40}_{-40}$ & $0.26^{+0.07}_{-0.08}$ & $0.99^{+0.02}_{-0.01}$ & $3^{+1}_{-1}$ & $0.8^{+0.1}_{-0.1}$($0.7^{+0.1}_{-0.1}$) & $240^{+10}_{-10}$($189^{+7}_{-11}$) \\  \\ 
\rowcolor{lightgray} $r_{s}=16\,\kpc$ & 86 & $470^{+40}_{-40}$ & $0.27^{+0.07}_{-0.07}$ & $0.99^{+0.02}_{-0.01}$ & $3^{+1}_{-1}$ & $0.8^{+0.1}_{-0.1}$($0.7^{+0.07}_{-0.12}$) & $240^{+10}_{-10}$($187^{+7}_{-10}$) \\  \\ 
$\sigma_g=800\,\kms$ & 86 & $470^{+40}_{-40}$ & $0.25^{+0.08}_{-0.07}$ & $0.98^{+0.03}_{-0.01}$ & $3^{+1}_{-1}$ & $0.8^{+0.1}_{-0.2}$($0.7^{+0.1}_{-0.1}$) & $240^{+10}_{-20}$($189^{+8}_{-11}$) \\  \\ 
\rowcolor{lightgray} $V_\odot+V_\mathrm{LSR}=240\,\kms$ & 86 & $480^{+40}_{-40}$ & $0.27^{+0.07}_{-0.07}$ & $0.99^{+0.02}_{-0.01}$ & $3^{+1}_{-1}$ & $0.8^{+0.1}_{-0.1}$($0.71^{+0.07}_{-0.13}$) & $240^{+10}_{-10}$($188^{+7}_{-11}$) \\  \\ 
\rowcolor{lightgray} $V_\odot+V_\mathrm{LSR}=260\,\kms$ & 86 & $470^{+40}_{-40}$ & $0.27^{+0.07}_{-0.08}$ & $0.99^{+0.02}_{-0.01}$ & $3^{+1}_{-1}$ & $0.8^{+0.1}_{-0.1}$($0.7^{+0.07}_{-0.12}$) & $240^{+10}_{-10}$($187^{+6}_{-10}$) \\  \\ 
\hline
\end{tabular} 
\medskip
\end{table*}

In this work, we present new measurements of the escape velocity curve $\vesc(r)$ and the virial properties of the M31 galaxy. Below, we first present these findings, and later discuss the possible systematics that may have crept into our final results due to some assumptions that we had to make.
Table~\ref{table:pneresult} summarizes the inferred values of our prime model parameters for multitude of cases. 

\subsection{Escape velocity: inference}
\begin{figure}
    \centering
	\includegraphics[width=1\columnwidth]{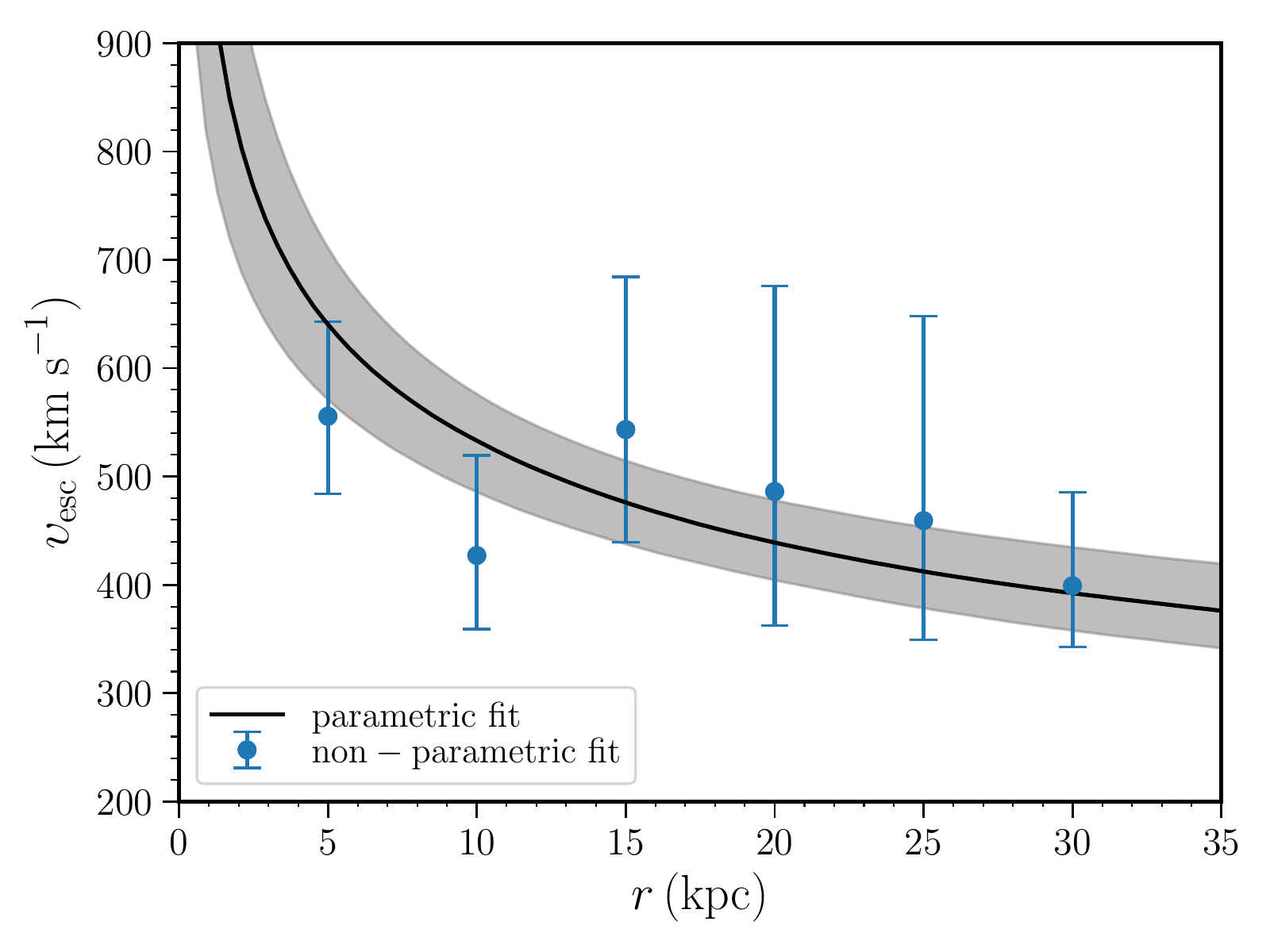}
    \caption{Estimated escape velocity curve using parametric (the black solid line with grey shades) and non-parametric (the blue dot with error bars) approaches with default assumptions (results corresponding to the first row in Table~\ref{table:pneresult}).}
    \label{fig:vesc}
\end{figure}
Fig.~\ref{fig:posteriors} shows the joint probability distributions of the model parameters obtained from a parametric fit of $\vesc$ (equation ~\ref{eqn:vescp}) to the high velocity PNe sample assuming `default' values of the parameters and key physical properties of the M31, that is, the threshold velocity $\vmin=300\,\kms$, $r_s=15\,\kpc$, $d_\mathrm{M31}=780\,\kpc$, $v_\mathrm{M31,h}=-301\,\kms$, $\sigma_g=1000\,\kms$, $\nu_g=1$ and uniform priors on $k$ and $\vesco$; the first row of Table~\ref{table:pneresult} presents the measurements for this case. 
Using our scheme, we derive an escape velocity curve of the M31 and its logarithmic power-law slope $\gamma=0.26\pm{0.07}$, and measure the escape velocity $\vesco=470\pm{40}\,\kms$ at the galacto-centric radius of $r_s=15\,\kpc$.
The parametric run of the $\vesc$ is shown with the black solid line in Fig.~\ref{fig:vesc}, where the grey band around the line shows the associated uncertainties inferred from the posterior distribution.
Complementing the parametric approach, we also present the $\vesc$ measurements at the given radii obtained from a non-parametric fit (Equation~\ref{eqn:vescnp}), which is shown in the figure with blue dots and vertical error bars.
We see that the non-parametric run has larger uncertainties in comparison to the parametric case, which is expected as the non-parametric case has larger degree of freedom due to the increased number of free parameters compared to the other case \footnote{known as the bias-variance trade-off \citep{eosl}}.
Comparing the results from the two different fits, in overall, we see that the results agree within the range of uncertainties and thus, validates our choice for the parametric expression of $\vesc(r)$ (i.e., equation~\ref{eqn:vescp}). 
However, there are some small systematic differences.
For $R<15$ kpc, the non-parametric estimate of $v_{\rm esc}$ is less than 
the parametric one, while for $R>15$ kpc, opposite is true. 
The observed H\,{\sc i} rotation curve also shows a similar behaviour with respect to our derived parametric rotation curve (Fig.~\ref{fig:vcirc}). This suggests 
that our parametric model is not flexible enough to capture the full information 
available in the data. Since the systematic deviations are small, 
we only consider the parametric form so as to take advantage of the reduced variance and modelling complexity.

Unfortunately, we are unable to constrain some model parameters that were kept free.
Importantly, we find that although a large value is preferred for $k$, it remains unconstrained. 
Unconstrained $k$ parameters due to scant number of high velocity stars have remained the main limitation of the undertaken approach, the fate that has been also shared by the MW high velocity works \citep[][refer to their figures 1 and 7 respectively]{1990ApJ...353..486L,2007MNRAS.379..755S}. 
Also, we find that the logarithmic number density slope favour shallow $(\alpha = 2)$ over steep model $(\alpha = 3)$ and the outlier fraction $\eta \to 1$ suggesting that effectively only a tiny fraction of our tracers are actually drawn from the background model. 

\subsection{Virial properties: inference}
Here, we first substitute the measured $\vesc(r)$ run into Equation~\ref{eqn:pot} to derive the total potential of the galaxy.
Then we fit the three component galaxy model to the derived potential and ultimately estimate the virial properties ($\mvir$ and $c$) of the dark matter halo, keeping the disk and bulge models fixed with Equation~\ref{eqn:bdparams}. 
This exercise is repeated separately for both the definitions of halo virial over-density.
In Fig.~\ref{fig:posteriors} inset we show the joint distributions of the $\mvir(M_{200})$ and $c$ resulted from this exercise in blue(red) contours for a case with `default' values of the model parameters. 
The marginalized distributions of the halo parameters are shown in the corresponding colours at the top and the right sides of the inset-figure.
Unfortunately, the marginalized distributions for the $c$ hint that within a large range of parameter space that we explore i.e. [1,60], we are unable to constrain the $c$ parameter, but we are able to measure the virial masses regardless; our final estimates for the $\mvir\,(M_{200})$ (also shown in the first row in Table~\ref{table:pneresult}) are $0.8\pm{0.1}\,(0.7\pm{0.1})\, \times \, 10^{12}\,\msun$.
Substituting these values for the masses in equation~\ref{eqn:rvir}, we determine $\rvir\,(r_{200}) = 240\pm{10}\,(188^{+7}_{-11})~\kpc$.
As discussed earlier in Section~\ref{sec:massmodel}, we note that the $\mvir$ or $\rvir$ values are always greater than for $M_{200}$ or $r_{200}$.

\subsection{Relaxing the assumptions}\label{sec:systematics}
Finally, it is instructive to investigate repercussions of relaxing some of the key assumptions on our final results, which we discuss below. 

\subsubsection*{Effects of change in M31 central properties}
Arguably, the heliocentric distance and the line-of-sight velocity of the M31 galaxy are known fairly accurately, with the scatter of roughly $1-2\%$ amongst the literature
\citep[e.g.,][]{1991rc3..book.....D,1998ApJ...503L.131S,2005MNRAS.356..979M,2008ApJ...678..187V,2012ApJ...758...11C}.
Nevertheless, we systematically shift the adopted value of $d_\mathrm{M31}=780\kpc$ and $\upsilon_\mathrm{M31}= -301\kms$ (Section~\ref{sec:coord}) by $\pm{10}\kpc$ and $\pm 5\kms$ respectively and re-run our analysis to understand their implications on our final results. 
Rows 2-5 in Table~\ref{table:pneresult} provide the estimates of the main model parameters for these cases. 
Importantly, we see that while the introduced systematic shifts in the $d_\mathrm{M31}$ have as such negligible effect on our final results, 
the biases introduced in the $\upsilon_\mathrm{M31}$ alters the $|\mvir|$ estimates by $\sim 8\%$.
Furthermore, one of our assumption is that the Sun's total tangential velocity relative to the Galactic centre, $V_\odot+V_\textrm{LSR}$ sums to $251.5\,\kms$ noting that it is approximately equal to the literature averaged value of $248\,\kms$ (for $R_\odot=8.2\kpc$) from \cite{2016ARA&A..54..529B}.
Since it is apparent from equations~\ref{eqn:vpec} and \ref{eqn:vpecT} that $\vpp \propto (V_\odot+V_\textrm{LSR})$, we test our choice.
For this we adopt two different values $240\,\kms$ and $260\,\kms$ for the Sun's total tangential velocity and re-run the analysis.
The measurements are reported in the bottom two rows of the Table~\ref{table:pneresult}, and it is evident that the systematic of $\pm{10}\kms$ in $V_\odot+V_\textrm{LSR}$ does not alter our final estimates, and likewise when we adopt slightly different values of $U_\odot=10\,\kms$ and $W_\odot=7\,\kms$ from \cite{2016ARA&A..54..529B}.
The null effect is expected because the $T$ term defined in equation~\ref{eqn:vpecT} that contains these constants, for a small angle $\Omega$, largely cancels out in equation~\ref{eqn:vpec}.

\subsubsection*{Effects of change in the threshold velocity ($\vmin$)}
As already described, $\vmin$ is the lower bound on the magnitude of the $\vpp$ that is used to discriminate the high velocity stars. 
In our final results, we adopt $\vmin=300\kms$, which is consistent with the value adopted in the literature \cite{1990ApJ...353..486L,2007MNRAS.379..755S,2014A&A...562A..91P}.
However, here we desire to investigate the effect of change in $\vmin$.
For this first we lower the limit and set $\vmin=290\,\kms$, this increases sample size to 111 as more stars are now classified as high velocity.
With the new set of data we find that the $\mvir/10^{12}\,\msun$ reduces marginally from 0.8 to 0.76.
Second, when we increase the limit and set $\vmin=310\,\kms$ we find that the sample size reduces to 65
and the $\mvir/10^{12}\,\msun$ slightly increases from 0.8 to 0.9.
Moreover, we find that relaxing the assumption on $\vmin$ by $\pm10\,\kms$ leads to insignificant changes in the estimates of other parameters, e.g., $\vesco, \gamma$ as well. 
Rows 6-7 in Table~\ref{table:pneresult} provide the estimates of the model parameters for these cases. 

\subsubsection*{Indiscernible effects due to change in some parameters}
There are also some model dependencies that we tested but which led to negligible changes in our final results, we briefly discuss them here.
Throughout the paper, to model the outliers (Equation~\ref{eqn:outlier}), we assume that the $\sigma_g=1000\kms$ and $\nu_g=1$.
We test that adopting $800\kms$ or $1200\kms$ for $\sigma_g$ and/or varying $\nu_g\in[0.5,2]$ effectively do not result any changes in the final $\mvir$ measurements.
Similarly, given the large distance between M31-MW ($d_\mathrm{M31}\gg R_0$) the assumed position of the Sun on the MW disk ($R_\odot=8.2\,\kpc$ versus $8.5\,\kpc$) has negligible effect in our analysis.
Moreover, we check that our final results are robust to the change of at-least $5\%$ in the upper limits of the prior ranges assumed for $\vesc, \gamma, \vpp^t$ and $\alpha$. 
Similarly, reducing the lower limit of prior range for $\alpha$ parameter from 2.0 to 1.0 do not modify our final $\mvir$ measurement.
A slight change of $\pm{1}\,\kpc$ in the adopted value of escape velocity scale-length $r_s = 15\,\kpc$ has no discernible effect in our mass measurement.
Results discussed in this Section are again provided in Table~\ref{table:pneresult}.

\section{Discussion}\label{sec:discussion}
\subsection{Independent validation}
It is important to affirm our mass model with existing independent measurements, which are based on completely different physics and hence, should have different systematics. In summary;

\subsubsection*{With the rotation curve}
\begin{figure}
    \centering
	\includegraphics[width=1.05\columnwidth]{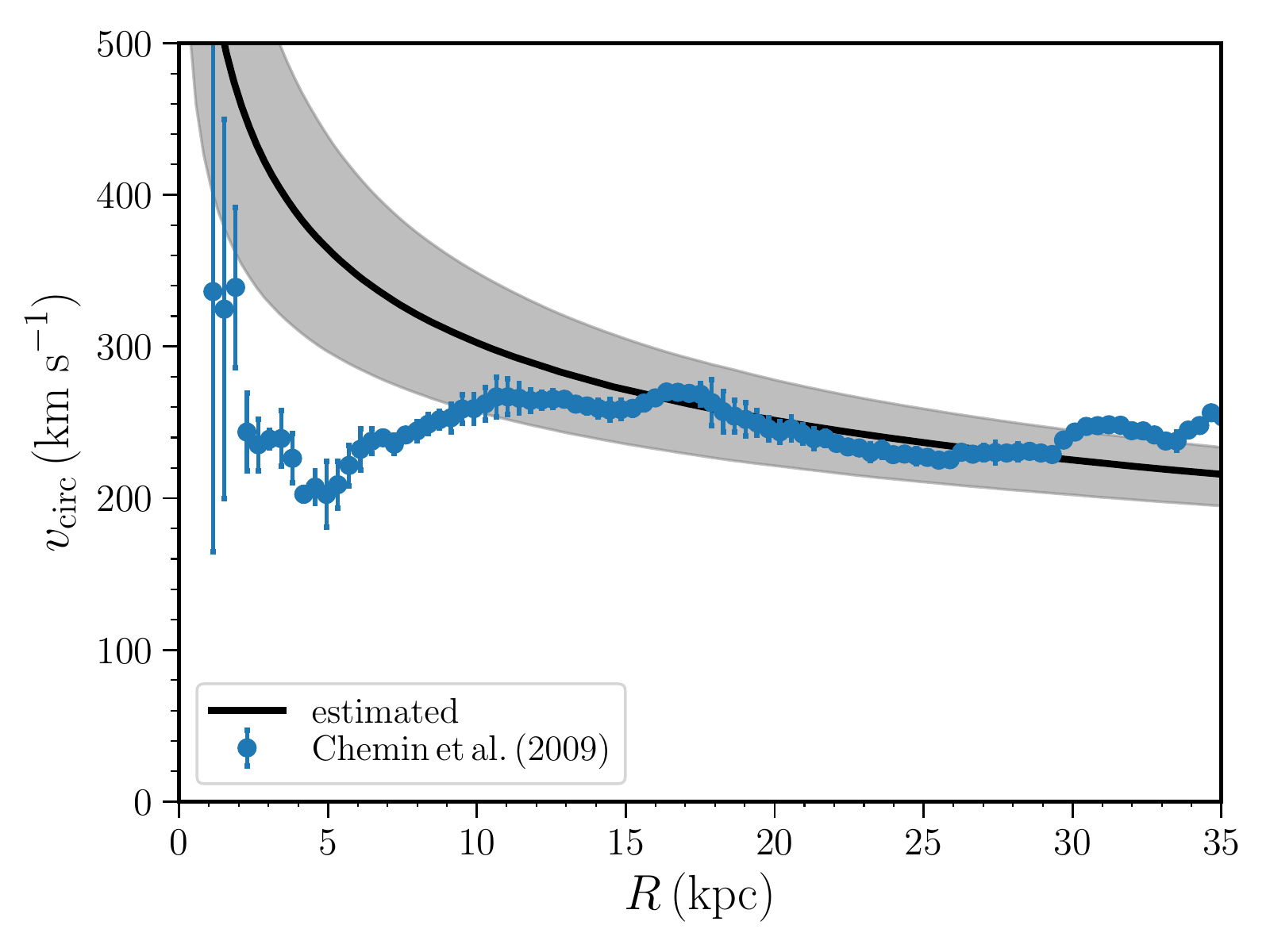}
    \caption{Circular velocity curve of the M31. Blue dots with error bars are measured values by \protect\cite{2009ApJ...705.1395C} using H\,{\sc i} emission line observations whereas grey shade with black solid line is our best estimate.}
    \label{fig:vcirc}
\end{figure}
Rotation curves have been derived for the M31 galaxy and the history of the measurement dates back to the seminal works by \cite{1970ApJ...159..379R,1980ApJ...238..471R} that use the optical data from H$\alpha$ emission lines.
Similarly, the curves have also been derived from the radio data using H\,{\sc i} or CO observations \citep[e.g.][]{1975ApJ...201..327R,1995A&A...301...68L,2006ApJ...641L.109C,2009ApJ...705.1395C,2010A&A...511A..89C}. 
It is crucial that we check our model prediction for the rotation curve with these independent measurements. 
In the spherical approximation and assuming that the gravitational and the centripetal acceleration are equal, we can derive the circular speed $\vcirc$ from the enclosed mass $M(<r)$ within radius $r$ using 
\begin{equation}
 \vcirc(r) = \frac{\text{G} M(<r)}{r},
\end{equation}
where the enclosed mass is derived from the escape velocity curve using the relation \[M(<r) = - \frac{r^2\, \vesc}{G} \frac{\mathrm{d}\vesc}{\mathrm{d}r},\]
that with Equation~\ref{eqn:vescp} it reduces to 
\begin{equation}\label{eqn:cumumass}
M(<r) = \frac{\gamma\,r}{\mathrm{G}} \vesc^2.
\end{equation}
The black line and the shaded grey regions in Fig.~\ref{fig:vcirc} show the derived $\vcirc$ and its associated uncertainty.
Additionally, the blue dots with error bars show an independent measurement of the M31 rotation curve, which is constructed using H\,{\sc i} emission line observations of five fields toward the galaxy; for details refer to \cite{2009ApJ...705.1395C}. 
Some notable features in the H\,{\sc i} rotation curve are that the velocities assume large values ($~350\kms$) in the innermost regions, which dips at $R=4\kpc$, gradually increases up to $\sim 270\,\kms$ at $\simeq15\kpc$ and remains flat ($230\,\kms$) out to $35\kpc$.
The $\vcirc$ we derive from the modelling of the high speed stars is largely in an agreement with the H\,{\sc i} rotation curve at $R\gtrsim10\,\kpc$, considering the large uncertainty around the $\vcirc$ estimate that we have.
Clearly, the two curves are not in agreement in the inner region, which is expected as the assumption of spherical symmetry of the disk used in deriving our $\vcirc$ profile.

We note that at $r_s=15\,\kpc$ the derived $\vesc$ is a factor of $\sim1.5$ times larger than $\sqrt{2}\vcirc$. If we look this result in conjunction with equation 14 of \cite{2007MNRAS.379..755S} or reference therein, interestingly, we can infer that there is a significant contribution to the $\vesc$ by mass exterior to $r_s$, demonstrating the presence of a dark halo in M31.

Finally, in Figure~\ref{fig:cumumass} we compare our derived cumulative mass profile of the M31 with different literature values and find that they are largely consistent. 
Exception to these are the M31 mass measurements using the tracer mass formalism \citep{2010MNRAS.406..264W,2014MNRAS.442.2929V}, which could be systematically off as the results are degenerate to the unknown velocity anisotropy parameter of the stellar halo. 

\subsubsection*{With the prediction of the baryonic Tully-Fisher relation (bTFR)} 
A fundamental empirical relationship between the total baryonic or stellar mass in galaxies and their maximum rotation velocity exists, which is known as the bTFR and it states that the baryonic mass of the spiral galaxy is proportional to the circular speed to the power of roughly $\sim4$ 
\citep{2000ApJ...533L..99M,2004PASA...21..412G,2009A&A...505..577T,2012AJ....143...40M,2012MNRAS.424.3123D,2015ApJ...802...18M}.
For our assumed value for the total stellar mass of the M31 of $13-20\times10^{10}\msun$ and adopting the proportionality constant of $41-53\, \msun/(\kms)^{4}$ from \cite{2012AJ....143...40M}, it is predicted that the range of maximum $\vcirc/\kms\in[222,264]$.
This estimate can be directly compared with the $\vcirc$ at the flat part of the measured rotation curve.
Consistent with the bTFR, we also see from Fig.~\ref{fig:vcirc} that our $\vcirc$ is equal to $250\,\kms$ say at $R\approx15\,\kpc$.

\subsubsection*{With the timing and momentum mass arguments}
As a final consistency check, we turn to the timing and mass arguments that utilize the relative motion of M31 and the MW to independently estimate the intergalactic mass or effectively the total mass of the Local Group.
From this, one can subtract off the MW mass, assuming it is known accurately enough, and directly predict the M31 mass.

The fact that M31 is moving radially toward the MW indicates that the mass of these galaxies must be sufficient to overcome cosmic expansion; this is known as the timing mass argument and was first proposed by \cite{1959ApJ...130..705K}.
Similarly, the argument that M31 and the MW should have equal and opposite momenta in the barycentric frame of reference of the Local Group is known as the momentum argument \citep{1982MNRAS.199...67E}.
Early works based on the timing mass argument, such as \cite{2008MNRAS.384.1459L,2012ApJ...753....8V}, have advocated for consistently high mass, i.e., $\sim5\times10^{12}\msun$, for the Local Group, 
suggesting the M31 mass should be roughly 4-6 times that of the MW.
However, more recently improved timing \citep{2014ApJ...793...91G} and momentum \citep{2014MNRAS.443.1688D} mass arguments have downward revised the M31-MW combined mass to $\sim2.5\pm{(0.4-1.5)}\times10^{12}\msun$. 
Our current estimate for the mass of M31, $0.8\pm{0.1}\times10^{12}\msun$, coupled with the MW mass of  $0.8-1.2\times10^{12}\msun$ \citep{2012ApJ...761...98K,2014ApJ...794...59K,2016ARA&A..54..529B}, implies a total mass of $\sim2\times10^{12}\msun$, which considering the uncertainties in both measurements lean supports to the downward revised mass of the MW-M31 system.

\subsection{Cosmological context}
It is essential to understand the properties of Local Group galaxies within a cosmological context.
Moreover, observations of the local universe have revealed that the existence of the `MW-Magellanic Clouds-M31'-like trio is cosmologically rare \citep{2012MNRAS.424.1448R}, which is even more scarce if we also seek for the planar configuration of the member satellites \citep{2015ApJ...805...67I}.
As our current theory of structure formation the \LCDM\ is getting ever more sophisticated mainly, due to increase in particle resolution, refinement of ``gastro-physics" and hydrodynamics, it is useful to compare its core predictions relevant to our observed results as it would allow us to establish the uniqueness or generality of the Local Group.
In the following, we compare our results against two profound theoretical predictions, that is, the concentration--virial mass scaling and the stellar-mass--halo-mass relations.

\subsubsection*{Comparison with concentration--virial mass relation}
\begin{figure*}
    \centering
	\includegraphics[width=0.91\columnwidth]{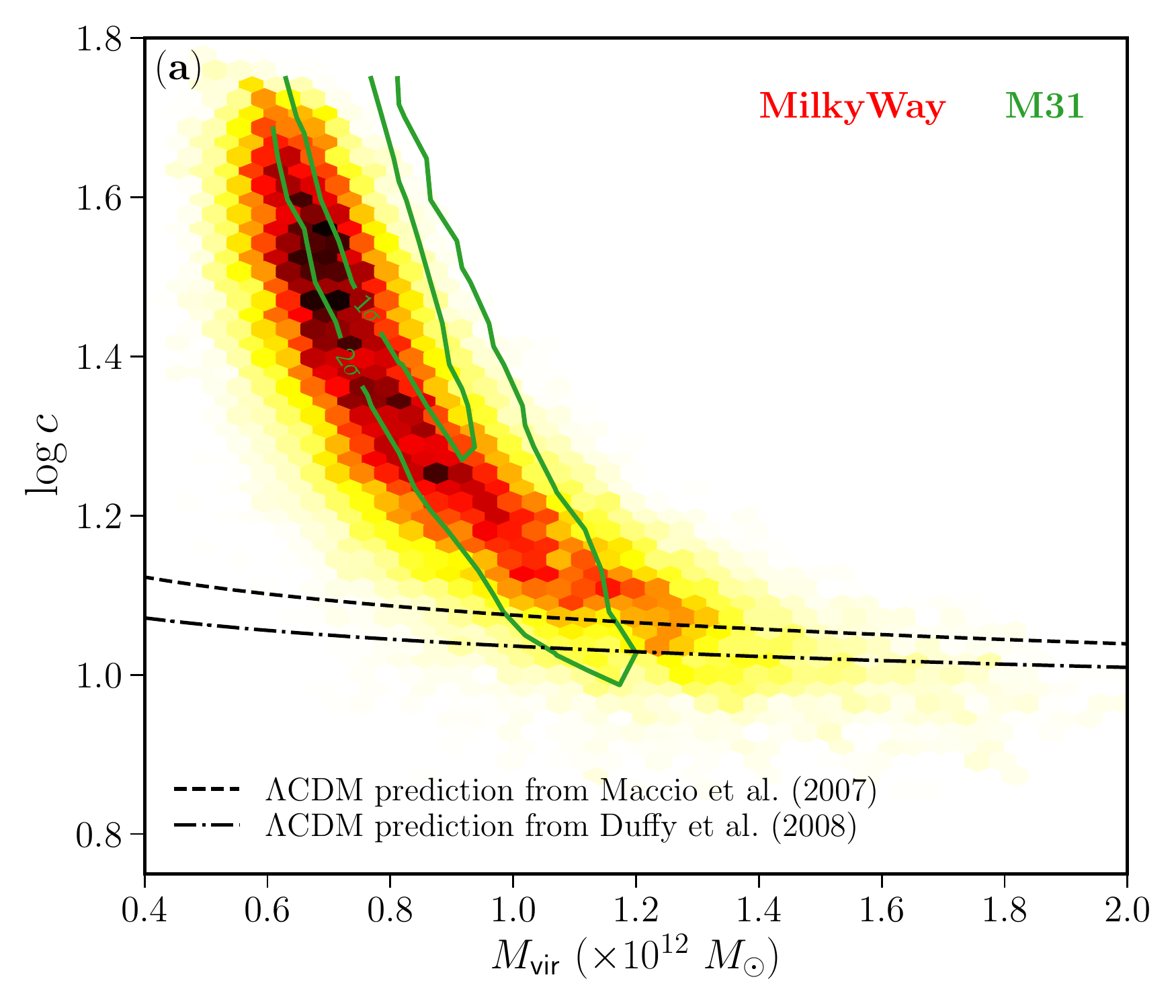}
    \includegraphics[width=1.05\columnwidth]{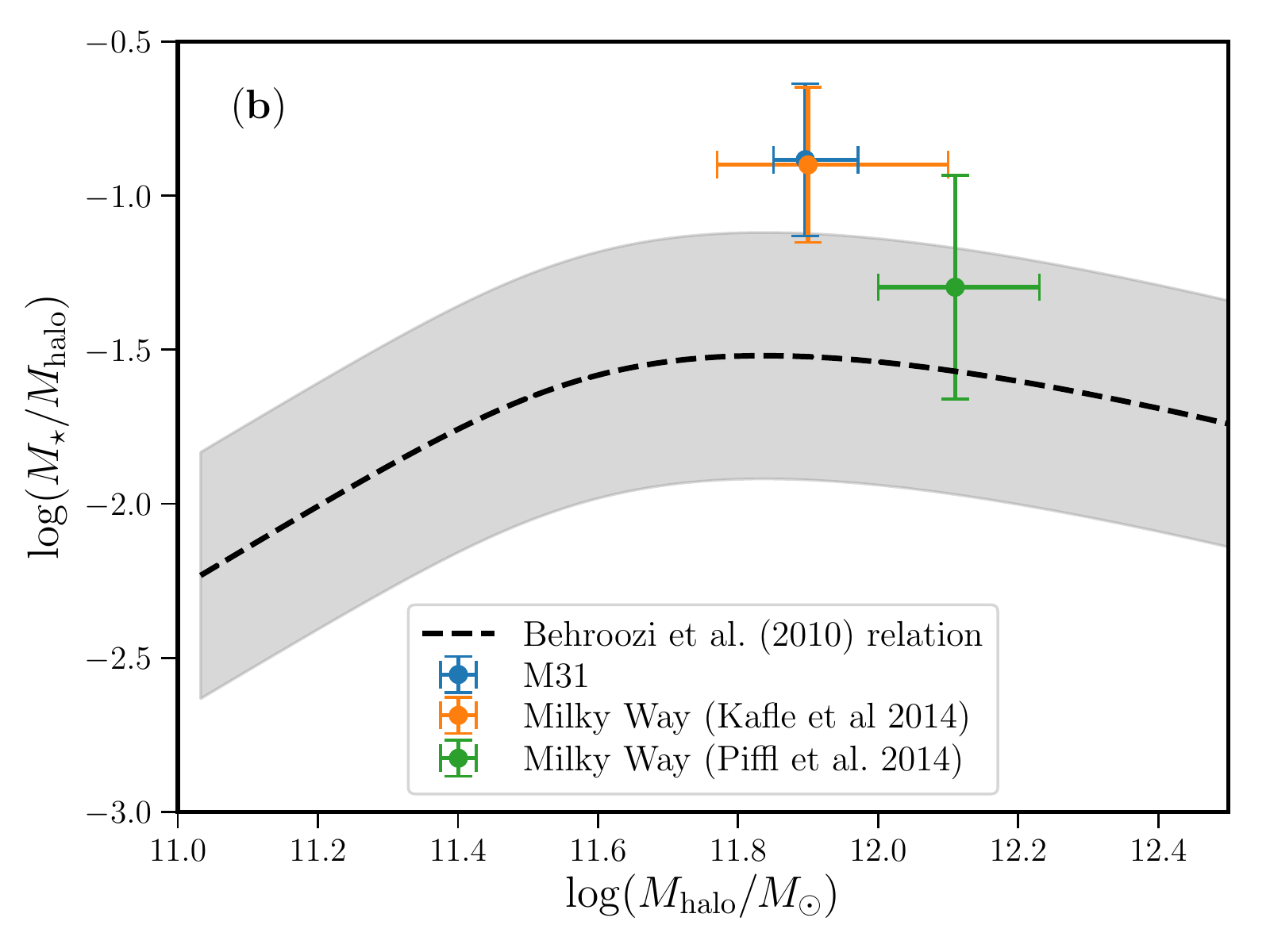}
    \caption{Comparison with the prediction of the \LCDM.
              Left panel (a) showing concentration ($c$)--virial mass ($\mvir$) joint probability distributions where green contour is our best estimate for the M31 whereas the heat map shows similar relation for the Milky Way taken from \protect\cite{2014ApJ...794...59K}. The black dashed and dash-dotted lines demonstrate a typical $c-\mvir$ relations predicted by \LCDM\ dark matter simulations of \protect\cite{2007MNRAS.378...55M,2008MNRAS.390L..64D} respectively.
Right panel (b): Stellar mass--halo mass relation expected from the abundance matching in the dark matter simulation, shown with the black dashed lines and grey shaded regions, where blue marker show the observed positions of the M31, and orange (from \protect\citealt{2014ApJ...794...59K}) and green (from \protect\citealt{2014A&A...562A..91P}) markers show the Milky Way galaxies respectively.}
    \label{fig:lcdmpred}
\end{figure*}
A generic prediction of simulated virialized halos in the \LCDM\ cosmology is that its mass and concentration are inversely proportional to each other, and assume a typical relation of $c \propto \mvir^{-0.12}$ for $\log(\mvir/\msun) \in [11,13]$ \citep[e.g.][etc.]{2001MNRAS.321..559B,2007MNRAS.378...55M,2008MNRAS.390L..64D,2011MNRAS.416.2539K,2016MNRAS.457.4340K,2016MNRAS.460.1214L}.
In Fig.~\ref{fig:lcdmpred}(a), we show the similar $c-\mvir$ relations from two representative fits from simulations \citet[][in black dashed line]{2007MNRAS.378...55M} and \citet[][in black dash-dotted line]{2008MNRAS.390L..64D} to which we overlap the green contours estimated of the M31 from Section~\ref{sec:results}. 
Similarly, the heat map shows the measured $c-\mvir$ relation for the MW from \cite{2014ApJ...794...59K}.
Interestingly, we see that both the galaxies show similar joint distributions although they are derived using different approaches and tracers, and barely overlaps with the \LCDM\ predictions at $2\sigma$ level.

\subsubsection*{Comparison with Stellar-Mass--Halo-Mass relation}
An another prediction of the \LCDM\ is that the stellar mass of the central galaxy embedded within a dark matter halo is correlated with its host halo mass, and is known as the stellar-mass--halo-mass relation \citep[SHM; e.g.][]{2010ApJ...710..903M,2010ApJ...717..379B,2013ApJ...770...57B}.
The SHM can be well defined by a double power law where on the either side of a peak-value the trends drop, and the peak-value corresponds to the knee of the stellar mass function \citep[e.g.][]{2012MNRAS.421..621B,2016MNRAS.457.1308M}.
In Fig.~\ref{fig:lcdmpred}b, the black dashed line shows a representative SHM adapted from \cite{2010ApJ...717..379B} near our range of interest, that is $\mvir\sim10^{11-12.5}$.
The grey shaded regions represent a typical scatter of $\sim0.3$ dex that the SHM relations in different literature sources are considered to have.
The blue dot with error bar marks our measurement for the halo mass of the M31 galaxy, given the disk and bulge combined stellar mass of $1.03\times10^{11}\,\msun$ taken from equation~\ref{eqn:bdparams}.
Clearly, the estimate for the M31 galaxy is higher than the generic prediction of the \LCDM.
Similarly, for a comparison we also show the green dot with error bar that represents the estimated values of dark matter halo ($1.3^{+0.4}_{-0.3}\times 10^{12}\,\msun$) and combined baryonic ($6.5\times10^{10}\msun$) masses of the MW adopted from \cite{2014A&A...562A..91P},
who uses identical modelling approach to this work. 
Additionally, in orange dot we show the yet another measurement of the MW mass, but from \cite{2014ApJ...794...59K} that uses the Jeans formalism.

Here, we like to leave a cautionary note that the above discussed theoretical predictions are for pure dark matter simulations, and are prone to serious systematics as they do not include baryonic processes such as cooling, star formation, and feedback. For example, the collapse of gas due to cooling leads to adiabatic contraction of the dark matter halo, which increases its concentration.
Feedback, on the other hand, can have the reverse effect. 
Similarly, the relationship between dark matter halos and galaxy stellar masses from the halo abundance matching technique rely on the accuracies of observed stellar mass function, the theoretical halo mass function and techniques of abundance matching.

Before we summarize our result, we point out some of the limitations of the escape velocity modelling technique. As highlighted in \cite{2014A&A...562A..91P} and \cite{2007MNRAS.379..755S} the conceptual underpinning of the technique that the 
density of stars in the velocity space tend to zero as velocity $\to \vesc$ (equation~\ref{eqn:df}) is fairly weak.
While in some analytic equilibrium models of stellar systems such as \cite{1983MNRAS.202..995J} and \cite{1990ApJ...356..359H} there is a non-zero probability density of finding a star all the way to the $\vesc$ and zero beyond that boundary; in other models, for example, the King-Michie  model \citep{1966AJ.....71...64K}, the density is zero even at speed $< \vesc$.
Moreover, it is unknown whether the phase space distribution of the galactic tracers extend to $\vesc$ and also whether it depends on the stellar types (for example, PNe versus regular stars). 
Any truncation in the velocity distribution will result an underestimation of the true $\vesc$, meaning in such case our estimates of $\vesc$ and thus $\mvir$ would only provide the lower limits of the true values. 
As such this will naturally solve the disagreement we note earlier between the theoretical stellar-mass--halo-mass relation and the observed value of the M31.
Being said that \cite{2007MNRAS.379..755S} notes that for a set of galaxies obtained from cosmological simulations \citep{2006MNRAS.365..747A} the level of truncation in the velocity distributions of the stellar component are found to be less than 10 per cent, and it seems not to have any systematic effect in their recovered $\vesc$.

\section{Summary}\label{sec:conclusion}
In this work, we establish a Hierarchical Bayesian framework in which we improve  the method by \cite{1990ApJ...353..486L}, and for the first time report an independent measurement of the radial dependence of the escape velocity curve, $\vesc(r)$, and a new estimate of the dynamical mass of the M31 galaxy.

We employ the Planetary Nebulae (PNe) data by \cite{2006MNRAS.369..120M} and \cite{2006MNRAS.369...97H} as it is currently the only publicly available tracer of the M31 disk with ample sample size and accurate enough line-of-sight velocity measurements.
After removing the marked extragalactic contaminants and satellite members, and stars with $|\vpp|\leqslant\vmin$, the final sample we analyse in our main results comprises 86 PNe.
From this high velocity sample of the PNe along-with our modelling scheme, we first estimate the $\vesc(r)$ and then derive the total galactic potential to which finally we fit a three component (bulge, disk and dark matter halo) mass model and infer the virial properties of the galaxy.
In the end, to provide a proper cosmological context, we discuss how the newly estimated mass of the M31 compares to the mass of our own the MW and also, to some generic predictions of the \LCDM\ such as the concentration-virial mass relation, the stellar mass--halo mass relation and the stellar Tully-Fisher relation. 

Following are the main conclusions of the paper:
\begin{enumerate}
 \item We present both the parametric and non-parametric $\vesc(r)$ profiles of the M31. Assuming the minimum threshold velocity $\vmin=300\kms$, the criteria set to classify high velocity stars, we measure $\vesc =470\pm{40}\kms$ at the galacto-centric radius of $r=15\kpc$. Additionally, we are also able to constrain the logarithmic power-law slope of the profile $\gamma=0.26\pm{0.07}$.
                     
 \item Using the derived $\vesc(r)$ profile and assuming spherical symmetry we are able to further derive the cumulative mass profile, the circular velocity profile as well as the total potential of the galaxy. To the derived potential we then fit a three component model of the galaxy --- Hernquist bulge, Miyamoto-Nagai disk and NFW dark matter halo model, of which we adopt the bulge and disk structural models from the literature \citep{2001ApJ...557L..39B,2012A&A...546A...4T} and keep them fixed. It is important to keep these components fixed because we only restrict our fitting to the derived potentials at $r\gtrsim10\,\kpc$ as we observe that only in this regime the derived circular velocity profile is in a good agreement with the rotation curve constructed from the H\,{\sc i} measurements. However, we keep the two defining parameters that is the concentration and the virial mass of the dark matter halo free. We find that assuming literature averaged bulge mass of $3.4 \times 10^{10}\,M_\odot$ and disk mass of $6.9 \times 10^{10}\,M_\odot$, the derived potential of the galaxy is best fit by a halo of the virial mass $\mvir(M_{200})=0.8\pm{0.1}\,(0.7\pm{0.1}) \times 10^{12} \msun$ that corresponds to the virial radius of $240\pm{10}\,(188^{+7}_{-11})\,\kpc$. 
 
 \item We find that the circular velocity curve ($\vcirc$) between $10\lesssim R/\kpc<35$ estimated by us is in good agreement with an independent prediction from the H\,{\sc i} observation. Similarly, the value of $\vcirc=250\,\kms$ we obtain at the flat part of the curve at $R\simeq15\kpc$ is consistent with the prediction from the baryonic Tully-Fisher relation. 
 We note that the measured $c-\mvir$ joint distributions and the observed locus of the stellar and dark matter halo masses of the M31 barely agree to the theoretical predictions at only $2\sigma$ levels.  
\end{enumerate}

\section*{Acknowledgements}
We like to thank the referee Prof. Mathias Steinmetz for a constructive and insightful report. 

{\it Software credit}: 
{\sc astropy} \citep{astropy}, {\sc bmcmc} \citep{2017ascl.soft09009S}, {\sc chainconsumer} \citep{chainconsumer}, {\sc daft} \url{https://github.com/dfm/daft},
{\sc ipython} \citep{ipython}, {\sc matplotlib} \citep{matplotlib}, {\sc numpy} \citep{numpy}, {\sc pandas} \citep{pandas} and {\sc scipy} \citep{scipy}.

\bibliographystyle{mnras}
\bibliography{paper.bbl}

\bsp	
\label{lastpage}
\end{document}